\documentclass[12pt]{article}
\usepackage{times}
\usepackage{marvosym}
\usepackage{geometry}
\usepackage{inputenc,graphicx}
\usepackage[most]{tcolorbox}
\usepackage{enumitem,amssymb}
\usepackage{ragged2e}
\usepackage{xcolor}
\newlist{thematic}{itemize}{8}
\setlist[thematic]{label=$\square$}
\usepackage{pifont}
\newcommand{\xmark}{\ding{55}}%

\begin{document}
\raggedright
\huge
The Importance of Telescope Training in Data Interpretation  \linebreak
\normalsize

\noindent \textbf{Thematic Areas:} \hspace*{60pt} $\square$ Ground
Based Project \hspace*{10pt} $\square$ Space Based
Project \hspace*{20pt}\linebreak $\square$ Infrastructure
Activity \hspace*{31pt} $\square$ Technological Development Activity
\linebreak \text{\rlap{\xmark}}$\square$ State of the Profession
Consideration \hspace*{65pt} \linebreak

\textbf{Principal Author:}

Name: David G. Whelan
 \linebreak						
Institution: Austin College
 \linebreak
Email: \texttt{dwhelan@austincollege.edu}
 \linebreak
Phone:  903-813-2345
 \linebreak
 
\textbf{Co-Authors:} (names, institutions, email) \linebreak
George C. Privon, University of Florida, \texttt{george.privon@ufl.edu} \linebreak
Rachael L. Beaton, Princeton University, \texttt{rbeaton@princeton.edu} \linebreak
Misty Bentz, Georgia State University, \texttt{bentz@astro.gsu.edu} \linebreak
S. Drew Chojnowski, New Mexico State University, \texttt{drewski@nmsu.edu} \linebreak
Jonathan Labadie-Bartz, University of S\~{a}o Paulo, \texttt{jbartz@usp.br} \linebreak
Gregory Mace, McDonald Observatory at UT Austin, \texttt{gmace@utexas.edu} \linebreak
Ryan Maderak, Benedictine College, \texttt{rmaderak@benedictine.edu} \linebreak
Steven R. Majewski, University of Virginia, \texttt{srm4n@virginia.edu}\linebreak
David Nidever, Montana State University, \texttt{david.nidever@montana.edu} \linebreak
James Webb, Florida International University, \texttt{webbj@fiu.edu} \linebreak


\justifying
\pagenumbering{gobble}
\pagebreak
\pagenumbering{arabic}

\section*{Overview}

Remote observing and the use of data obtained from mid- to large-scale
astronomical surveys now accounts for a sizable portion of the data
used for research in astronomy. This shift has enabled novel research
activities and has democratized them to the extent of making large
datasets available to anyone with an internet connection. As a side
effect, fewer astronomers are being trained on the basics of telescope
operation. The dangers associated with this reduced training include
the risk that professional astronomers lose the ability to discern
instrumental from astrophysical signals in their data. Additionally,
there is a real risk that technical skills that were once associated
with professional practitioners will be relegated to the realm of
support staff and amateurs. This may result in the average population
of professional astronomers possessing inferior technical skills in
observing, and therefore less creativity and fewer insights in the
pursuit of their research.

In this State of the Profession Consideration, we will discuss the
state of hands-on observing within the profession, including:
information about professional observing trends; student telescope
training, beginning at the undergraduate and graduate levels, as a key
to ensuring a base level of technical understanding among astronomers;
the role that amateurs can take moving forward; the impact of
telescope training on using survey data effectively; and the need for
modest investments in new, standard instrumentation at mid-size
aperture telescope facilities to ensure their usefulness for the next
decade.

\begin{tcolorbox}[left=0mm,right=5mm,width=6.5in,colback=gray!5!white,colframe=gray!75!black]
\textbf{\large Recommendations:}
\justify
\begin{itemize}
\item Professional observing facilities allocate funds to support
  travel for observers who wish to observe on-site, with favor shown
  to junior researchers (including students and early post-doctoral
  fellows)
\item We encourage both university and national stakeholders (e.g.,
  the National Science Foundation) to support the development of new
  campus observing facilities and the upgrade and maintenance of
  existing facilities
\item Mid-size (2- to 5-meter class) national, state, and consortium
  facilities operating as traditional observatories be provided with
  funds to upgrade their standard instrumentation suite, to ensure
  their usefulness for the next decade
\end{itemize}
\end{tcolorbox}

\pagebreak

\section{Modern Observational Astronomy}\label{sec:modern}

Large-area surveys of the sky and observing queues for space- and
ground-based observatories are innovations made possible by computing
power and the internet. Since the Infrared Astronomical Satellite
(IRAS) came out with a full-sky map and fluxes for every detected
source, we have increasingly depended upon such surveys, and now the
great majority of astronomers use results from surveys as part of
their normal practice.

Surveys have allowed astronomers to study very large samples at
once. This has been revolutionary, when one considers how difficult it
was to obtain data before the advent of electronic detectors; in the
mid- to late-twentieth century sample sizes of hundreds of objects
were considered quite large, whereas thousands to hundreds of
thousands of data points are more common today. The statistical
results from these new, large surveys have greatly increased our
understanding of things such as the mass-metallicity relation for
galaxies (e.g., Tremonti et al. 2004), Baryon Acoustic Oscillations
(e.g., Tojeiro et al. 2014) and stellar population studies (e.g.,
Gouliermis et al. 2006). They have become so common that we accept as
common language discussions concerning statistical likelihood of model
fits to our large samples. And the growing use of Artifical
Intelligence in the search for data-driven solutions that are of a
statistical nature, are regularly used to conduct research.

The ready availability of big data means that trips to traditional
telescopes to collect what are now considered small samples of data
are becoming increasingly rare. Some facilities have shut their doors;
others have been threatened with closure\footnote{e.g., at Kitt Peak
  \url{www.mira.org/newsletr/NL\_spring14web.pdf}}; still others have
transitioned from traditional observing mode to survey\footnote{the
  Dark Energy Spectroscopic Instrument (DESI) will be the sole
  instrument on the Mayall 4-meter telescope on Kitt Peak:
  \url{https://www.desi.lbl.gov/telescope/}}. And psychologically,
this only seems to re-enforce the greater value of survey data, or the
quick follow-up that is possible with queue observing, over
traditional observing runs. And yet much important work in astronomy
is being conducted traditionally, including: calculations of the
primordial Helium abundance (e.g., Izotov et al. 2014), hunting for
young solar analogs (Gray et al. 2015), and follow-up observations of
transient events (e.g., emission-line B-type stars; Labadie-Bartz et
al. 2018, and exoplanets; Huber et al. 2019).

A close look at the big data trend suggests that the astronomical
community is at risk of losing something vital, however. Without
visting telescopes to collect data, astronomers are one step removed
from considering fundamental issues such as focus and image quality,
signs of optical aberrations and how to mitigate their effects, the
quality of the acquired flatfield images and whether superflats will
need to be constructed, or the shape of the instrument profile in
spectroscopy and whether it can reasonably be removed. Allowing
computers to handle so many of the mundane aspects of practical
observing such as those listed here sets the dangerous precedent of
de-valuing the base knowledge that may not be in itself publishable.

Do we de-value the most fundamental knowledge about observing? It
would seem not: staff astronomers at observing facilities are
generally tasked with ensuring that the data product, beginning in the
raw form and processed by whatever automatic data reduction and
extraction pipeline has been created, is of the highest possible
quality. For surveys, studies are performed to assess data quality and
derive uncertainties in quoted results. These studies are done to
provide the astronomical community with necessary information about
the limits of the dataset and any appropriate caveats. It is up to
members of the community who wish to use these data to read the
appropriate documentation in order to educate themselves about these
limitations.

But what about observers using survey and queue data products? How can
we offer opportunities for them to gain the lower, more fundamental
knowledge necessary to interpret their data? Out of respect for the
scientific method, is it possible to question the validity of the
automatic pipeline reduction without offending the staff astronomers
who have worked diligently to create the data products we wish to use?
The problem is that automatic pipelines risk serving as black boxes,
about which little is known and no questions may be asked. And this
level of trust {\it is} unscientific; since the people responsible for
making the automatic pipelines are fallible, not every mistake can be
caught by them. And since mistakes may lurk in the reduced data
product, then it is possible (and has in fact happened) that observers
could treat an instrumental signature for an astronomical one, or
else, by doing the opposite and ignoring small details as spurious,
lose valuable insight that is available to them, if only they knew how
to look.

\section{The Continued Need for Hands-on Observing Training}\label{sec:required}

As described above, mere exposure to pipeline-reduced data products or
data secured by queue observing does not suffice to teach observers
about the importance of the lower-level fundamentals of their
trade. We contend that the only real method available to deepen
understanding about telescopes and telescope data is active
participation in data collection, including trouble-shooting problems
on-sky, and reducing one's own data according to the accepted
methods. This presents a difficulty, in that trips to observing
facilities are expensive, and funding is not always available as it
once was to enable them. There is the additional consideration of how
to be conscious about our carbon footprint. But we are here discussing
the need for training, not that queue and remote observing should be
done away with; in fact, queue and remote observing modes are
excellent tools that allow us to be more environmentally
conscientious.

Our first recommendation to foster telescope training is that
\textbf{observing facilities be allowed to allocate funds to support
  travel for observers who wish to observe on-site}. This type of
support is particularly important for junior researchers (from
students to early post-doctoral fellows), who may be doing on-site
training for the first time, or have a vested interest in securing
data for their future careers.

Travel may be prohibitive for a number of reasons, and we therefore
\textbf{urge members of the community at colleges and universities to
  investigate avenues by which to support their own campus
  observatories}. Studies have shown (Benn \& S\'{a}nchez 2001,2005;
Trimble et al. 2005) that telescopes in the 1-meter class routinely
produce cutting-edge data products, outperforming expectations based
on studies comparing publication and citation rates {\it vs.} aperture
size. This is clearly indicative of the great number of unsolved
questions lurking among the bright and nearby in the sky. Campus
observatories also have two practical advantages over larger, national
facilities for the purposes of training, in that they are close to the
students, and, by virtue of being relatively small, offer observers
the opportunity to learn about their telescope, instruments, and
environment in great detail. Put simply, campus observatories offer
the single best way to teach future astronomers the fundamental
observing practices, by providing ample time on sky, and room for
making mistakes without using time at high-cost professional
observatories.

We believe so strongly in the use of college and university
observatories to train observers that we additionally recommend that
\textbf{national funds, perhaps through the National Science
  Foundation, be made available for the creation of new campus
  facilities and the refurbishment of existing, out-of-date ones.}
Something like a {\it Minor} Research Instrumentation Program would be
a low-cost opportunity with the singular goals of increasing
accessibility to observing facilities for students: it is possible to
make an entire facility for something on the order to \$250,000 or
less (including telescope, mount, instrumentation, and software), and
for purchases related to refurbishment (CCD cameras and commercial
spectrographs) to total less than \$25,000.

It is reasonable to question whether college and university
observatories will, before or after being funded, be used in the
absence of a willing professional astronomer to oversee it. We have
learned from personal experience that interested members of a faculty
with limited prior experience with telescopes can easily be trained
starting at an amateur level, largely because of the amount of
material available to amateur astronomers on the internet. When these
resources are found wanting, as they will if interest continues, one
often has colleagues elsewhere to call upon for assistance.

Furthermore, there are a number of projects already in existence that
are perfect for college observatories to partake in, particularly when
starting a new facility means that there is pressure to demonstrate
its scientific usefulness. For exoplanets, the Kilo-Degree Extremely
Little Telescope (KELT; Pepper et al. 2007) has developed the KELT
Follow-Up Network, whereby small observatories can contribute to
follow-up observations, to discern real exoplanet transits from
eclipsing binaries or other phenomena. The Transiting Exoplanet Survey
Satellite (TESS) has a similar program\footnote{TESS-FOP:
  \url{https://tess.mit.edu/wp-content/uploads/TFOPcharterDWLv10.pdf}}.
Studies of classical emission-line B-type stars have the Be Star
Spectra (BeSS) Database\footnote{\url{http://basebe.obspm.fr/basebe/}}
to contribute to, studies of cataclysmic variables have the Center for
Backyard Astrophysics\footnote{\url{https://cbastro.org}}, and a
variety of other projects exist through the American Association of
Variable Star Observers\footnote{AAVSO:
  \url{https://www.aavso.org/observing-main}}. A local amateur
community may also be a well of enthusiasm and practical expertise.

For mid-size (2- to 5-meter class) professional observatories, there
is generally an issue that instrumentation is aging\footnote{See APC
  White Paper: ``The Importance of 4m Class Observatories to
  Astrophysics in the 2020s''(Chanover et al.)}. Funds may not be
available to make replacements, or there may be a lack of interest in
doing so, or else there may be a lack of vision in how this is to be
done. We believe that professional observatories deserve to be
furnished with a suite of up-to-date instrumentation, and recommend at
least the following:

\begin{itemize}
\item a large-format CCD camera and associated suite of astronomical
  filters for photometry;
\item a low- to mid-resolution spectrograph, in the optical and/or
  near-infrared; and
\item a high-resolution spectrograph, at similar wavelengths.
\end{itemize}

We strongly recommend that \textbf{mid-size (2- to 5-meter class)
  national, state, and consortium facilities operating as traditional
  observatories be provided with funds (federal, state, or private
  respectively) to update every facility to this minimum
  standard}. This instrumentation is the most common for follow-up
observations of transient events, and their standardized availability
will not only make these telescopes useful for the next decade, but
provide ample opportunity for collaboration between observers from
different sites\footnote{These items are complimentary with the
  motivations and recommendations described in the APC White Paper
  ``2020 Vision: Towards a Sustainable OIR System'' (Oey et al.)}. We
realize that, with funding as tight as it is these days, it may seem
redundant to produce the same suite of instruments at every mid-size
observatory; when this is seen as impractical, it might be beneficial
to \textbf{promote time trades between facilities with different and
  complementary instruments}. For telescope facilities with other
specialities at their disposal besides those listed above, such as a
Coud\'{e} feed, additional funding should be made available to update
the appropriate instrumentation, if there is a demonstrable need for
it from within the community.

\section{In Conclusion}\label{sec:conclusion}

The prevalence of large surveys of data within the astronomical
community risks conditioning us to rely too heavily on data products
created by a small percentage of people within our community. This
situation has endangered our ability to understand the true
limitations of the datasets that we work with -- a situation that can
only be corrected by actively seeking opportunities to take part in
the data collection process. To that end, we have made several
recommendations:

\begin{enumerate}
\item Funds be made available to support travel to telescopes,
  especially in aid of junior-level research;
\item Funds be made available to support the refurbishment or creation
  of campus observatories to train students and create research
  opportunities; and
\item Mid-size facilities take steps to modernize their
  instrumentation in a standard way, and/or promote time trades to
  increase their user base
\end{enumerate}

We additionally support ongoing efforts by members of faculties at
colleges and universities to create local opportunities for their
students.

Understanding that skills require upkeep, the ability of observational
astronomers today to analyze their data requires that they be properly
trained in the fundamentals of observing. This is what principally
allows us as a community to continue to make progress in unveiling the
secrets of the Universe.

\section{References}

\begin{enumerate}

\item Benn, C.~R., \& S\'{a}nchez, S.~F. 2001, PASP, 113, 385.

\item Benn, C.~R., \& S\'{a}nchez, S.~F. In {\it Communicating
  Astronomy}, T. Mahoney, ed. La Laguna, Tenerife, Spain: Instituto de
  Astrofisica de Canarias (IAC), 2005, pp.8-13.

\item Gouliermis, D.~A., Dolphin, A.~E., Brandner, W., \& Henning,
  Th. 2006, ApJS, 166, 549.

\item Gray, R.~O., Saken, J.~M., Corbally, C.~J., Briley, M.~M.,
  Lambert, R.~A., Fuller, V.~A., Newsome, I.~M., Seeds, M.~F., \&
  Kahvaz, Y. 2015, AJ, 150, 203.

\item Huber, D., et al. 2019, AJ, 157, 245.

\item Izotov, Y.~I., Thuan, T.~X., \& Guseva, N.~G. 2014, MNRAS, 445,
  778.

\item Labadie-Bartz, J., Chojnowski, S.~D., Whelan, D.~G., Pepper, J.;
  McSwain, M.~V., Borges Fernandes, M., Wisniewski, J.~P.,
  Stringfellow, G.~S., Carciofi, A.~C., Siverd, R.~J., Glazier, A.~L.;
  Anderson, S.~G., Caravello, A.~J., Stassun, K.~G., Lund, M.~B.,
  Stevens, D.~J., Rodriguez, J.~E., James, D.~J., \& Kuhn, R.~B. 2018,
  AJ, 155, 53.

\item Pepper, J., Pogge, R.~W., DePoy, D.~L., Marshall, J.~L., Stanek,
  K.~Z., Stutz, A.~M., Poindexter, S., Siverd, R., O'Brien, T.~P.,
  Trueblood, M., \& Trueblood, P. 2007, PASP, 119, 923.

\item Tojeiro, R., et al. 2014, MNRAS, 440, 2222.

\item Tremonti, C.~A., Heckman, T.~M., Kauffmann, G., Brinchmann, J.,
  Charlot, S., White, S.~D.~M., Seibert, M., Peng, E.~W., Schlegel,
  D.~J., Uomoto, A., Fukugita, M., \& Brinkmann, J. 2004, ApJ, 613,
  898.

\item Trimble, V., Zaich, P., \& Bosler, T. 2005, PASP, 117, 111.

\end{enumerate}

\end{document}